\documentclass[rnote]{aa}

\usepackage[varg]{txfonts}
\usepackage[amssymb]{SIunits}
\usepackage{graphicx}
\bibpunct{(}{)}{;}{a}{}{,} 

\begin{document}

%
%

\title{Fundamental and harmonic plasma emission in different plasma environments}
\titlerunning{Fundamental and harmonic plasma emission}

\author{U. Ganse\inst{1} \and
	P. Kilian\inst{2} \and F. Spanier\inst{3}
\and R. Vainio\inst{1,4}}

\offprints{U. Ganse, \email{urs.ganse@helsinki.fi}}

\institute{Department of Physics, University of Helsinki, 00014 Helsinki, Finland
	 \and 
	 Max-Planck Institut f\"ur Sonnensystemforschung, 37077 G\"ottingen, Germany
	 \and
	 Center for Space Research, North-West University, 2520 Potchefstrom, South Africa
	\and
	Department of Physics and Astronomy, University of Turku, 20014 Turku, Finland
}
	
\date{\today}

%
%

\abstract {} { 
Emission of radio waves from plasmas through plasma emission with fundamental
and harmonic frequencies is a familiar process known from solar type II radio bursts.
Current models assume the existence of counterstreaming electron beam
populations excited at shocks as sources for these emission features, which
limits the plasma parameters to reasonable heliospheric shock conditions.
However, situations in which counterstreaming electron beams are present can
also occur with different plasma parameters, such as higher magnetisation,
including but not limited to our Sun.
Similar radio emissions might also occur from these situations.
} { 
We used particle-in-cell simulations, to compare plasma microphysics of
radio emission processes from counterstreaming beams in different plasma
environments that differed in density and magnetization.
} { 
Although large differences in wave populations are evident, the emission process
of type II bursts appears to be qualitatively unaffected and shows the same
behaviour in all environments.
} {}


\maketitle

%
%

\section{Introduction}
Solar type II radio bursts are intermittent radio phenomena within the
heliosphere that have been observed and recorded since the earliest days of
astronomical radio observations \citep{WildMcCready}.
These bursts, observed in conjunction with coronal mass ejections (CME)
\citep{caneErickson} and shocks driven by other flare-related eruptions
\citep{StereoData, TypeIIFlare}, show a characteristic multibanded spectral
morphology: typically two frequency bands, called the fundamental and harmonic
emission, exist at frequencies typically between 10 and
\unit{1000}{\mega\hertz} and are drifting towards lower
frequencies at a rate of about \unit{0.25}{\mega\hertz\per\second} \citep{NelsonMelrose}.

\begin{figure}[htb]
	\begin{center}
		\includegraphics[width=\hsize]{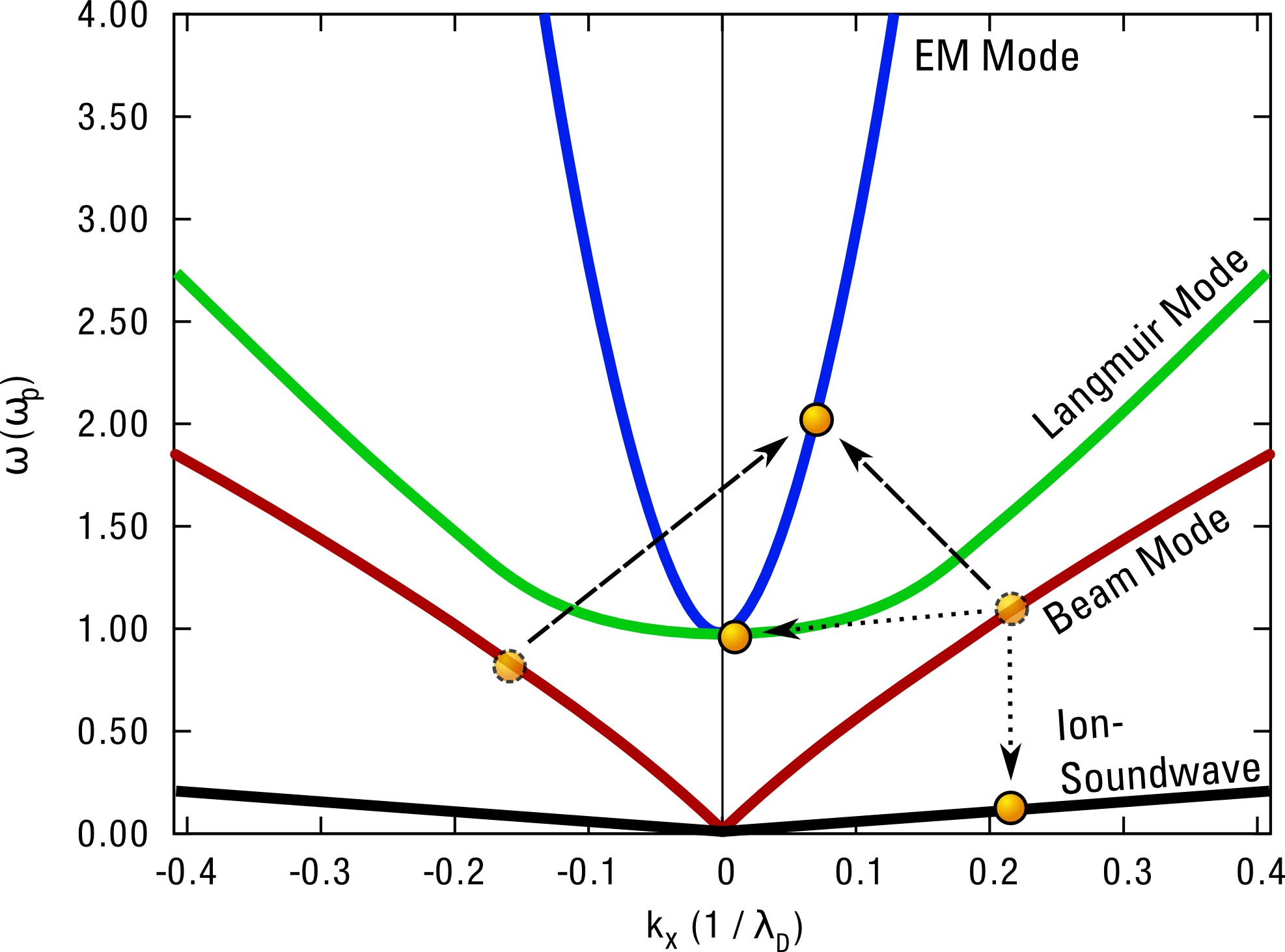}
	\end{center}
	\caption{Kinematics of the nonlinear wave coupling processes leading to type
	II radio burst emission: decay of electrostatic wave energy (dotted arrows) excites
	ion soundwaves and radio emissions at the plasma frequency $\omega_p$, while
	coalescence of counter-propagating electrostatic waves (dashed arrows)
	creates radio emissions at $2 \omega_p$}.
	\label{fig:dingsmerged}
\end{figure}

The emission process of these bursts arises from non-thermal
electron beam populations excited through shock drift acceleration at the
leading edge of the corresponding CME- or flare-driven shock
\citep{NelsonMelrose, WorkingGroupD, KnockModel}, causing beam-driven instabilities
\citep{MikhailovskiiCherenkov} in the foreshock region,
especially in the presence of counterstreaming electron beams caused by shock
ripples \citep{knockShockRipples}.
The resulting
electrostatic wavemodes can participate in nonlinear interaction processes, thus radiating
radio emission at the plasma frequency and its harmonic \citep{Melrose1970, Melrose, WillenGeneralizedLangmuir, SpanierVainio09}:
\begin{eqnarray}
	L &\rightarrow& S + T(\omega_p)
	\label{melroseGleichungen1}\\
	L + L' &\rightarrow& T(2 \omega_p).
	\label{melroseGleichungen2}
\end{eqnarray}
Here, $L$ and $L'$ denote forward- and backward-propagating beam-driven
Langmuir waves, $S$ denotes ion soundwaves and $T$ denotes transverse
electromagnetic modes.  These processes are depicted in Figure
\ref{fig:dingsmerged}.

Remote verification of the nonlinear interaction processes through radio
observations is hampered by the fact that nonlinear plasma excitations can only
leave the emission regions where they couple to propagating linear modes - in
this case, where the nonlinear excitations are coincident with the
electromagnetic modes' dispersion relation. These are detectable as the
familiar fundamental and harmonic frequency bands of type II radio bursts,
which do not, by themselves, carry any information about the $k$-space
distribution of the nonlinear processes that originally excited them.

Even in scarce in situ spacecraft observations of radio burst emission regions
\citep{Pulupa2007}, only partial information about the wave populations present
there are obtainable, since the pointwise measurement of wave spectrometers
can only provide $\omega$ but no $k$ information. Only in computer
simulations can complete spatiotemporal information on any wave quantity be obtained.

In previous works, the authors have investigated the processes in this emission
region using a particle-in-cell simulation model \citep{acronym11,
HockneyEastwood}: the dependence of the emission intensity on the beam strength
has been probed by \citet{GanseApj2012} and
uni- versus bidirectional electron beams have been compared by
\citet{ursSolarPhysics}.

Because they are directly produced via nonlinear plasma processes near a coronal
shock front \citep{knockShockRipples},
the plasma environments are constrained by conditions under which
shock-formation is possible -- specifically, the magnetic field strength and
consequently the Alfv\'en velocity has to be low enough for super-Alfv\'enic
velocities to be achieved by solar ejecta. Thus, an effective upper limit for
the Alfv\'en velocity at shocks of about $\unit{2000}{\kilo\meter\per\second}$
exists.
In a CME's loop structure, however, no such limitation exists and
counterstreaming beam situations can potentially occur in much stronger
magnetic fields, caused for example by magnetic bottle configurations.
These environments are not constrained to small spatial extents and
heliospheric-shock magnetic field strengths, so emissions from these regions
can be assumed to be broadband in nature.
In a recent study by \citet{PohjolainenWideband}, a number of wideband type II
radio burst events are shown to propagate slower than the shock they are
associated with, which might be another indication that the emission process
is not necessarily confined to the foreshock region and that a continuum of
phenomena between type II and type IV bursts may exist.

In addition to at the Sun, radio bursts have also been observed for nearby flare stars
\citep{AreciboObservation}, where higher densities and magnetic fields are also implied.

The open question in this model is whether the three-wave interaction processes
confirmed to take place at low magnetic field strengths can equally contribute
to radio emissions from areas where magnetic fields are significantly stronger,
or whether synchrotron radiation is the only viable emission mechanism.

Whereas theoretical treatment of type II bursts \citep{Melrose} is typically
based on the assumption of only one electromagnetic mode being present in the
emitting medium, splitting into R- and L-mode (for field-parallel propagation)
or X- and O-mode (for propagation perpendicular to the magnetic field) can
potentially alter the interaction behaviour.

Therefore, by varying the simulation setup in plasma density and magnetisation,
the viability of the same emission process for broadband type II/IV radio bursts
is tested here.

\section*{Setup}

\begin{table}
	\caption{Simulation parameters of the three particle-in-cell simulation runs.}
	\label{fig:tab1}
	\centering
	\begin{tabular}{crrrrr}
		\hline
		\hline
		\textbf{\#} & \textbf{Simulation size}/cells & $\vec{B_0}$/G & $\unit{\rho_e}{\per\centi\meter^{-3}}$ & $\Omega_{ce} / \omega_{pe}$\\ 
		\hline
		1 & $8192 \times 4096$ &  $1$ & $1{.}25\cdot10^{7}$ & 0.093\\
		2 & $4096 \times 4096$ & ${0{.}015}$ & $1{.}05\cdot10^2$ & 0.689\\
		3 & $4096 \times 4096$ &   ${0{.}22}$ & $3{.}06\cdot10^{3}$ & 1.729\\
		\hline
		\hline
	\end{tabular}
\end{table}
Simulations were performed using the ACRONYM particle-in-cell code
\citep{acronym11}, a fully relativistic electromagnetic code tuned for the
study of kinetic-scale plasma wave phenomena and -interactions.
The simulation setup was adopted from the previous works
\citep{GanseApj2012, ursSolarPhysics}, with a spatially periodic,
2.5-dimensional simulation box filled with a homogeneous plasma, into which two
counterstreaming non-thermal beam components were injected at $v_{\text{beam}}
\approx 5 \, v_{\text{th}}$.

Simulation \#1 used parameters based on observations of foreshock plasmas, with
weak magnetization and coronal density. For simulation \#2 and \#3, the ratio of
electron gyro frequency to plasma frequency was subsequently increased as
denoted in Table \ref{fig:tab1}, thus representing plasma environments with
correspondingly stronger magnetization. (It should be noted that both plasma
density and magnetic field strength are actually decreasing in the simulation
parameters. However, the particle-in-cell simulation method allows arbitrary rescaling
with respect to plasma frequency, so the results are valid for all plasmas of
the given frequency ratio.)

The corresponding change in Debye length, gyroradius, and plasma frequency
resulted in appropriately larger physical extents of the simulation, which were
all run at roughly the same numerical resolution.
We note that the particle distribution in velocity space remained
unchanged between all three runs, meaning that all simulations were run with
the same thermal and electron beam velocities.

\section*{Results}

\begin{figure*}[htbp]
	\begin{center}
		\includegraphics[width=\hsize]{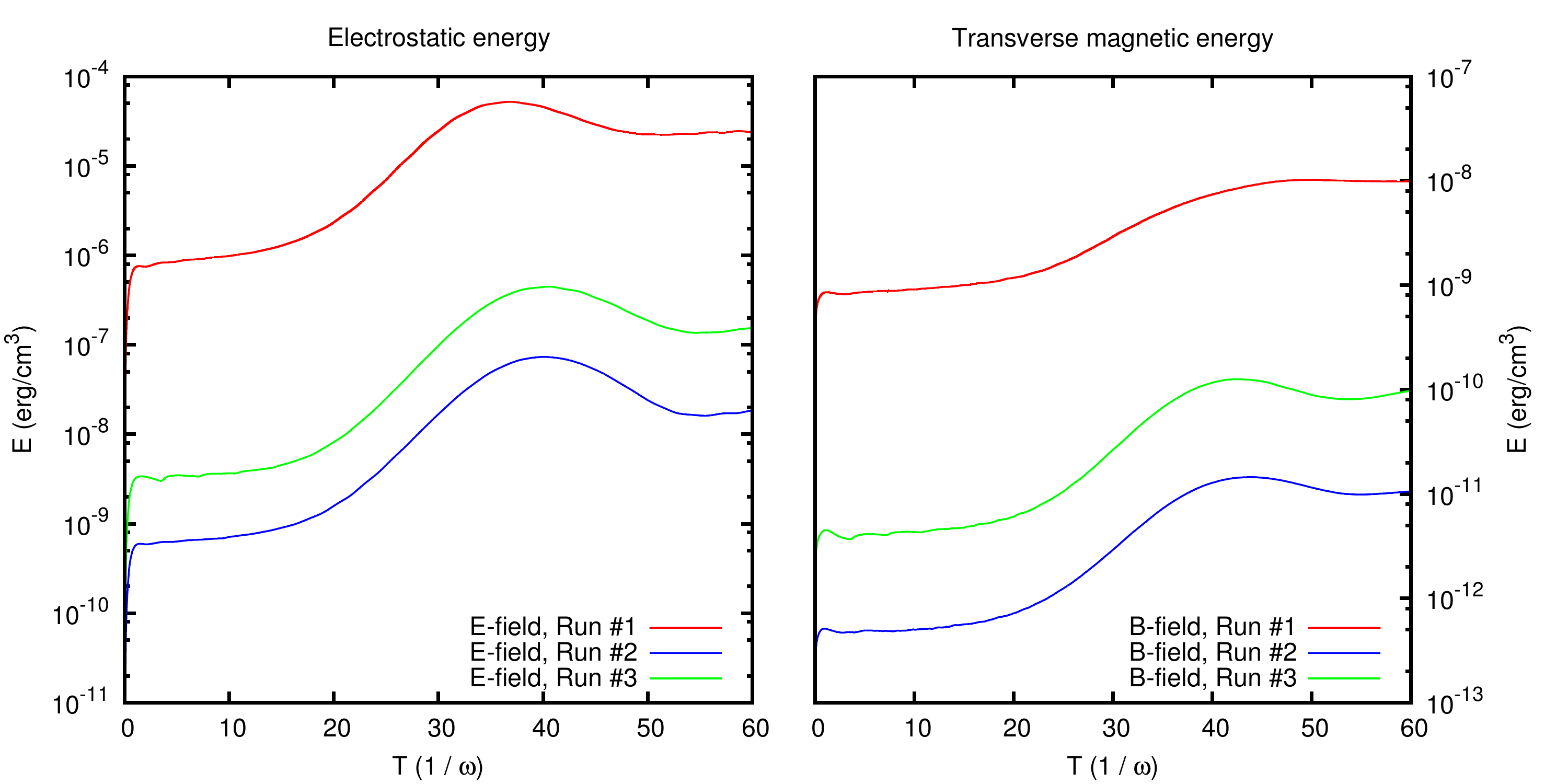}
	\end{center}
	\caption{Distribution of electric field (as a measure of electrostatic wave
	intensity) and transverse magnetic field strength (as a measure of electromagnetic
	wave intensity) in simulation runs \#1, \#2 and \#3, shown
	in red, blue and green.}
	\label{fig:distanceenergyouts}
\end{figure*}

To investigate the predicted nonlinear wave couplings, both electrostatic and
electromagnetic wave behaviour have to be analyzed. The simplest quantity for
assessing their respective intensities is the total energy content of
longitudinal electric fields (as a measurement of electrostatic waves) and the
transverse magnetic field energy (as a measurement of magnetic field
intensity), integrated over the complete simulation box. Figure
\ref{fig:distanceenergyouts} shows the development of these quantities over the
time span of the simulation runs.  The electric field energy initially rises
strongly, as beam-driven wave excitation forms the electrostatic wave
population. This process eventually
saturates, as sufficient kinetic energy of the beam electrons is
depleted. Correspondingly, the transverse magnetic energy shows a much
slower start of its growth, consistent with the nonlinear excitation from
electrostatic wavemode couplings.

It is apparent that this behaviour is represented identically in all three
simulation runs, with their relative energy densities scaled by the changing
plasma density and spatial extent of the simulations.

\begin{figure}[htbp]
	\begin{center}
		\includegraphics[width=\hsize]{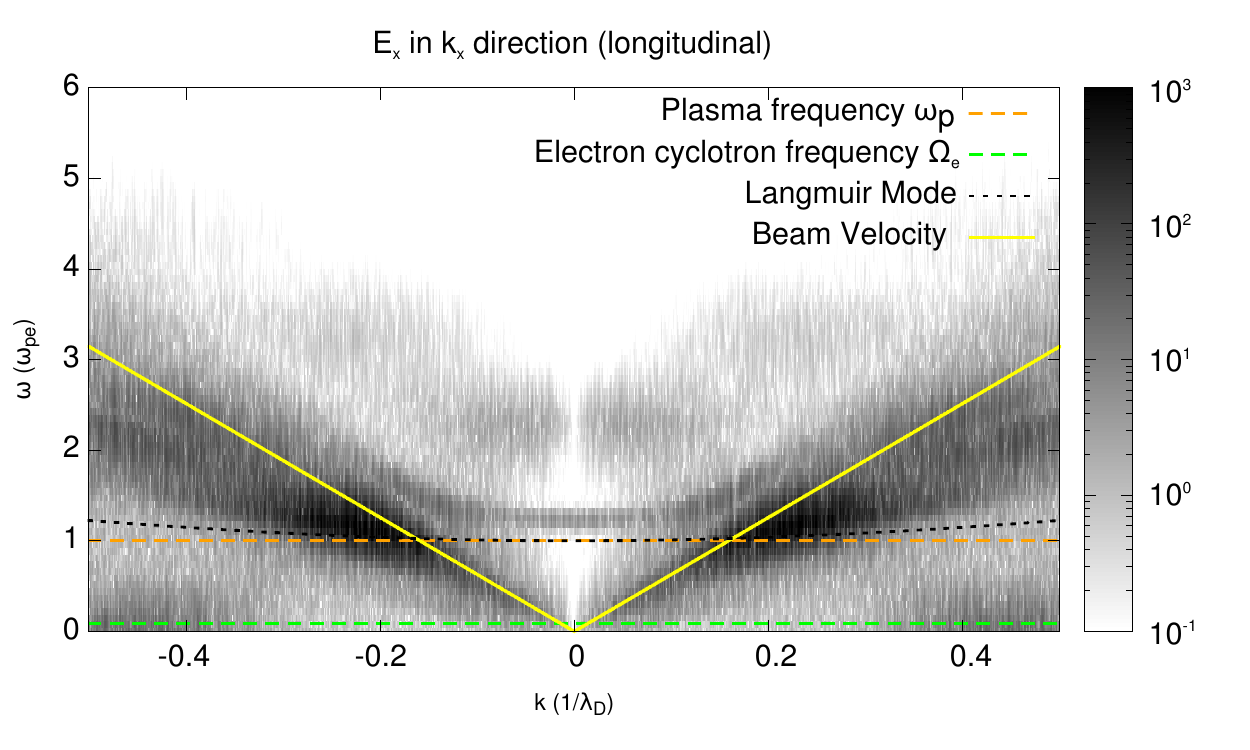}
		\includegraphics[width=\hsize]{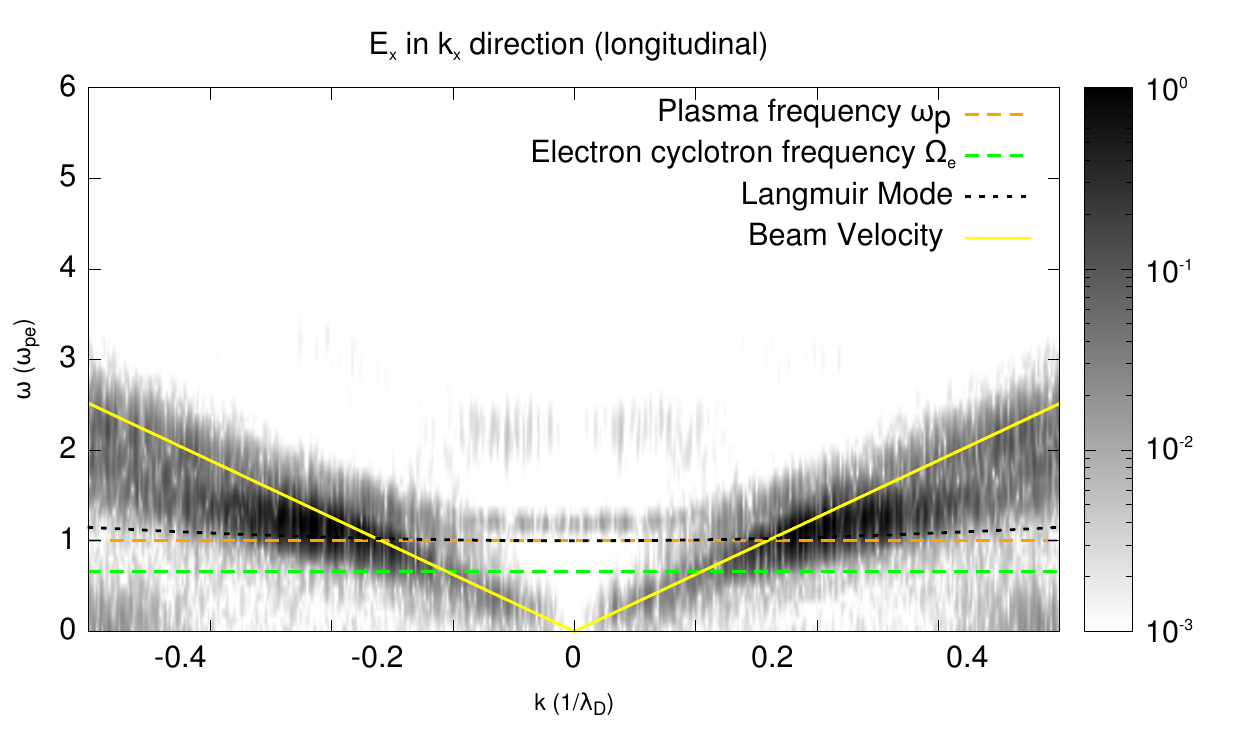}
		\includegraphics[width=\hsize]{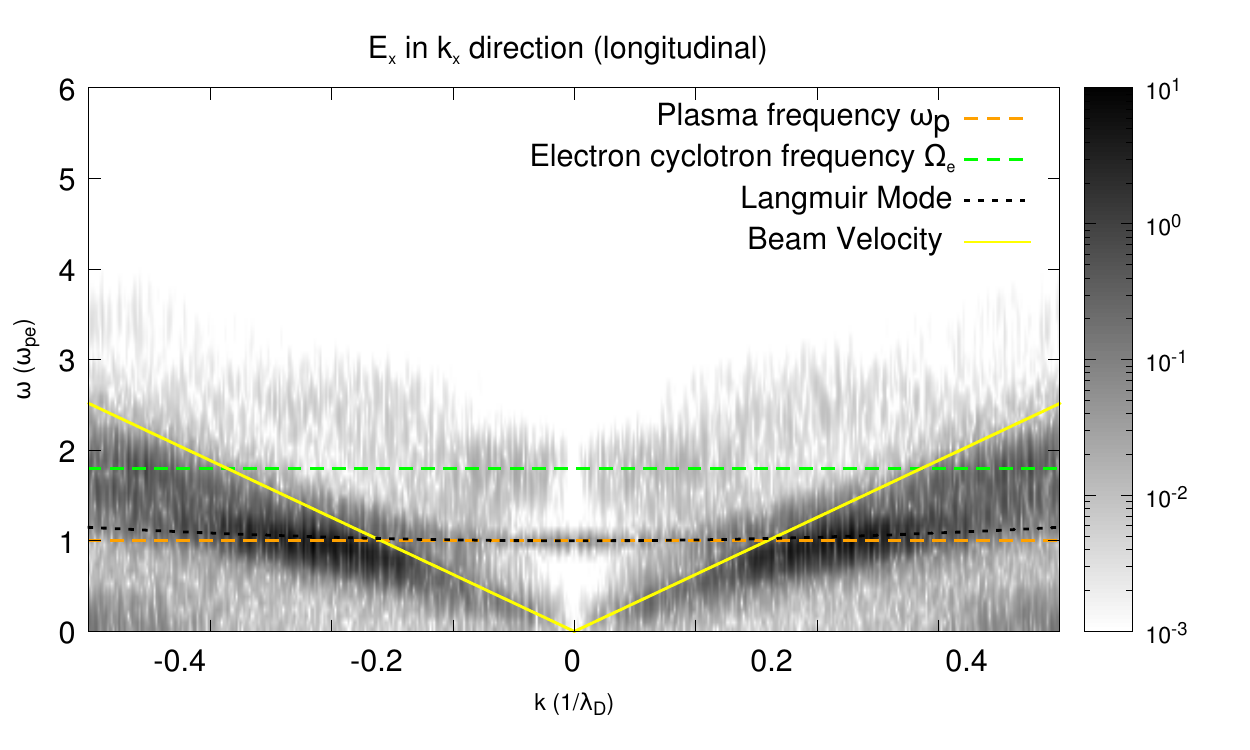}
	\end{center}
	\caption{Comparison of electrostatic wave dispersion and intensities for simulations with increasing magnetization.
	Apart from intensity rescaling due to different plasma densities, no morphological changes in wave behaviour occur.}
	\label{fig:edisps}
\end{figure}
For a more detailed comparison of the individual wavemodes, the field
quantities within the simulation were spatially and temporally
Fourier-transformed to obtain intensities in $k-\omega$ space, which allows for
direct identification of wavemodes through their characteristic dispersion
relations.
Figure \ref{fig:edisps} shows the dispersion behaviour of the longitudinal
electric field so obtained for each of the three simulation runs. A total
change in intensity due to plasma parameter rescaling based on density
again is the only change visible in these plots -- no morphological change of wave
excitation can be observed. The dominant wave feature in all three plots is
the beam-driven wave, whose dispersion forms a linear feature with a slope
corresponding to the electron beam velocity. The strongest excitation occurs
in resonance with the electron plasma frequency $\omega_{p}$, as predicted by
\citet{MikhailovskiiCherenkov} and \citet{WillenGeneralizedLangmuir}.

\begin{figure}[htbp]
	\begin{center}
		\includegraphics[width=\hsize]{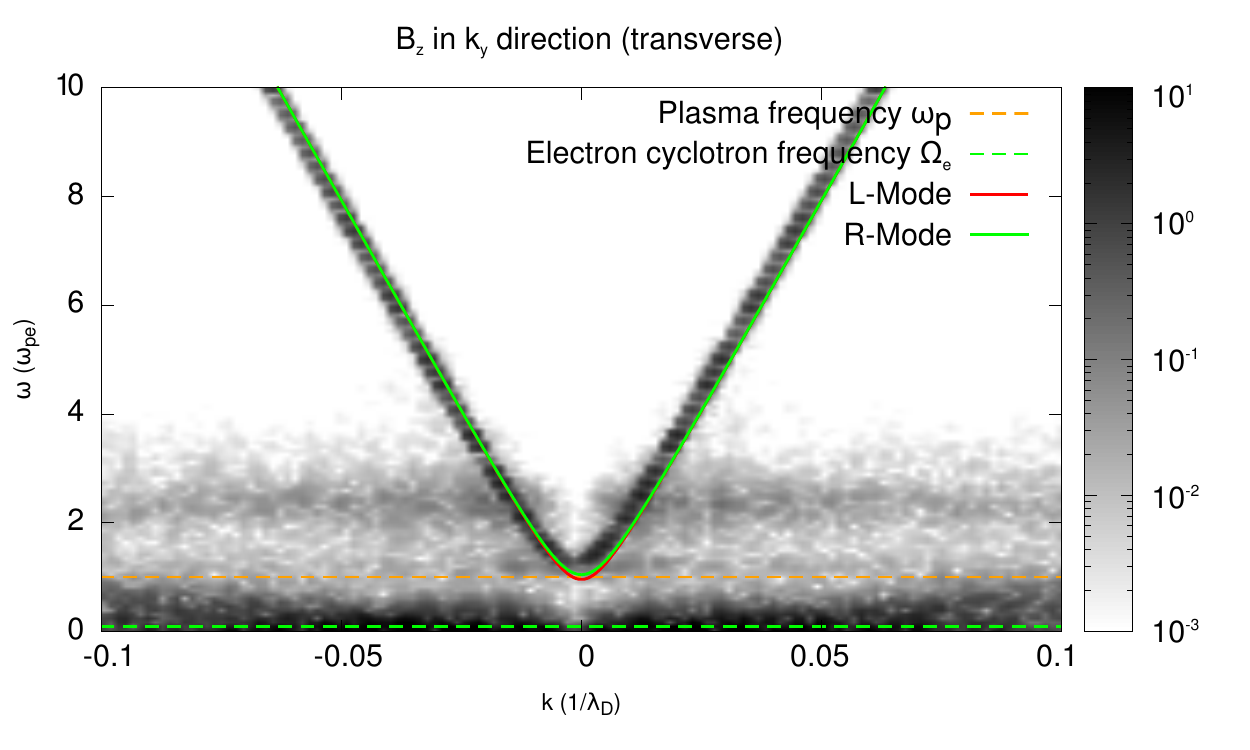}
		\includegraphics[width=\hsize]{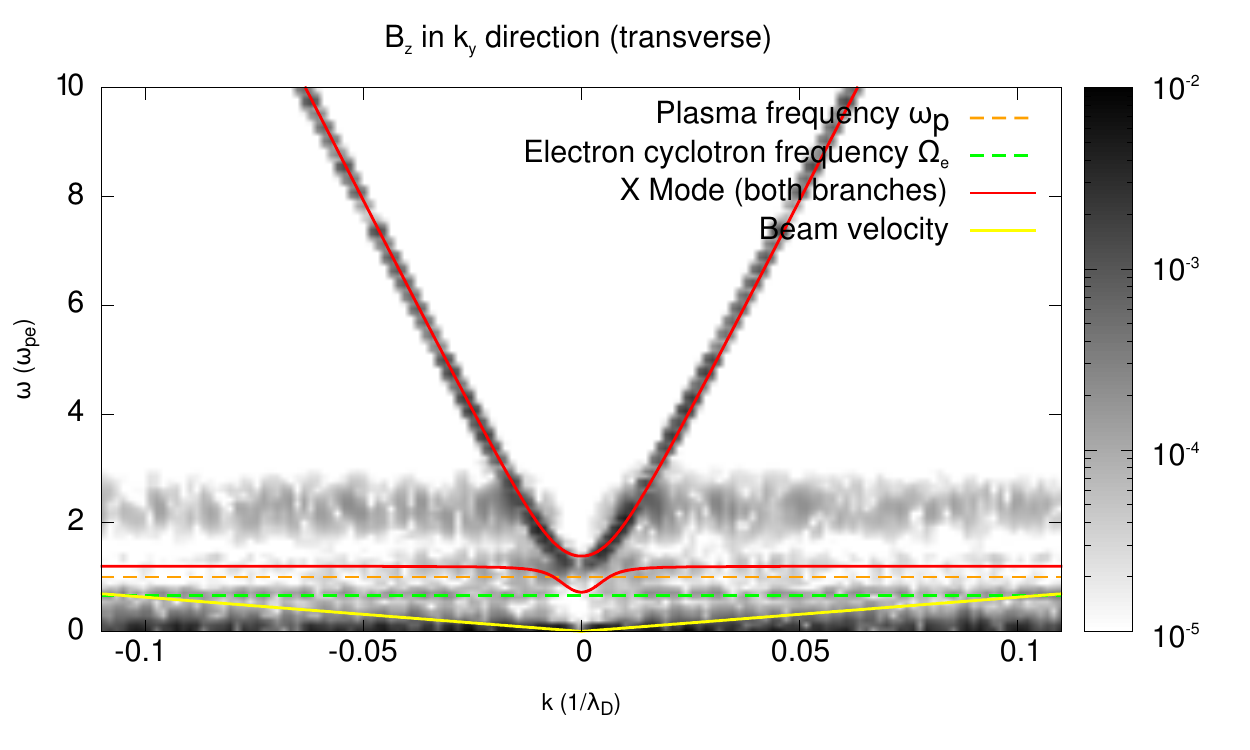}
		\includegraphics[width=\hsize]{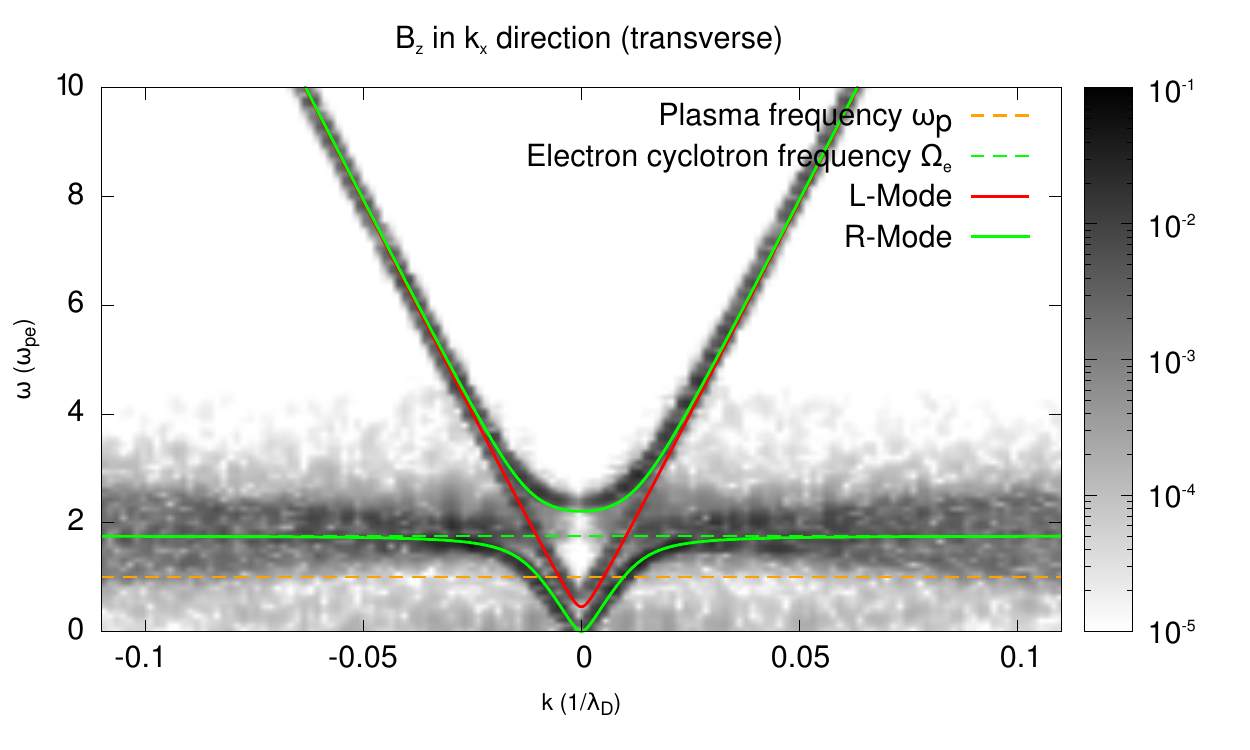}
	\end{center}
	\caption{Comparison of electromagnetic wave dispersion and intensities for
		simulations with different magnetization strengths: \#1 (top) \#2 (centre)
		and \#3 (bottom). While the spectral structure of electromagnetic modes
		varies strongly between the three runs, the horizontal bands of nonlinear
		wave excitation persist regardless of magnetic field strength.}
	\label{fig:bdisps}
\end{figure}

More interestingly, the dispersion behaviour of a transverse magnetic field
component in all three simulations is given in Figure \ref{fig:bdisps}, which
show the spatial and temporal Fourier transform of a transverse component of
the magnetic field. Transform direction was chosen along the background field
(for simulations \#1 and \#3) or orthogonal to it (for simulation \#2).
Here, the effect of changing magnetisation is clearly visible in the changed
mode composition between the three simulations: while simulation \#1
shows an electromagnetic mode that is nearly undisturbed by the
background magnetic field, the splitting into X- and O-mode is becoming
apparent in simulation \#2 and very pronounced splitting into R- and L-mode
in case \#3.
Still, common to all three simulation runs are horizontal emission bands at the
fundamental ($\omega_p$) and harmonic ($2 \omega_p$) frequency, which do not
correspond to any linear wave mode of the plasma. These features have
previously been identified \citep{GanseApj2012, KarlickyIAUS} as the
predicted nonlinear excitations caused by interaction of beam-driven modes and
Langmuir waves.
Observationally, these bands correspond to the radio excitations at $\omega_p$
and $2\omega_p$, identified as the fundamental and harmonic frequency bands of
type II radio bursts.

Because they are a product of electrostatic waves, these features appear to be largely
unaffected by the change in magnetisation, and are consistently occurring in
all plasma environments investigated here.
By coupling to the electromagnetic mode, in whatever configuration it may be
present within the emission region, the nonlinear process can thus lead to
radio emission at the fundamental and harmonic frequency.

To further investigate the development of energies contained in the fundamental
and harmonic emission bands, the latter half ($t>20/\omega_p$) of the
simulation was divided into a number of temporal subdivisions, and dispersion analysis was
run individually for each of the slices.
Because time and frequency are conjugated quantities, a reduction in temporal extent
leads to a reduction in frequency resolution in these dispersion snapshots. Hence,
as a compromise between sufficient temporal and spectral resolution, the times
were chosen such that both fundamental and harmonic emission bands remained
clearly discernable in the plots, leading to five data points for simulation \#1
and \#3, and three data points for simulation \#2.

\begin{figure}[htbp]
	\begin{center}
		\includegraphics[width=\hsize]{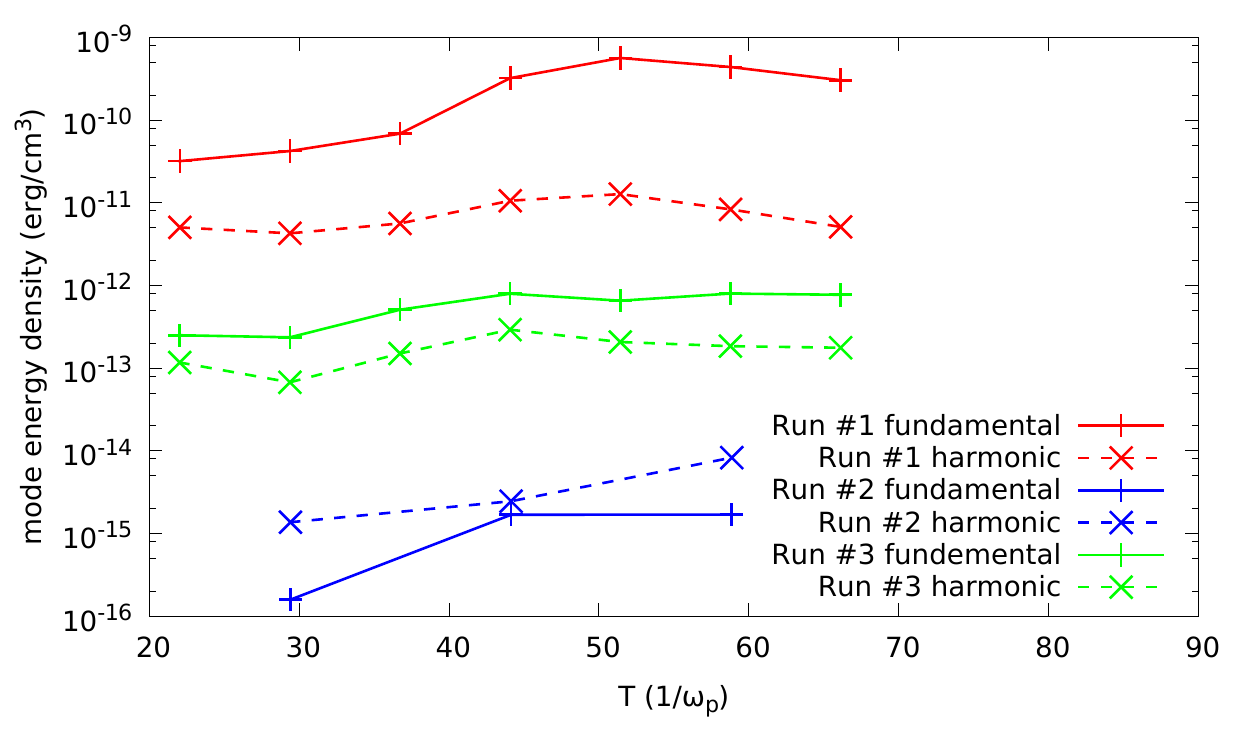}
	\end{center}
	\caption{Development of harmonic and fundamental mode energies in the second
	half of the simulation (where a sufficiently strong signal was discernible).}
	\label{fig:modeenergies}
\end{figure}

By summing over the intensities contained in these bands, we determined the mode intensity
development. Figure \ref{fig:modeenergies} shows the resulting
individual behaviour in the latter half of the simulations. In all three
cases, both modes follow the total energy development of the transverse magnetic fields.
While simulations \#1 and \#3 show more energy in the
fundamental than in the harmonic emission band, the reverse is true for
simulation \#2. The close proximity of the electron cyclotron frequency to the
harmonic band's frequency is assumed to be the culprit here -- in the dispersion
diagrams, the lack of a clear separation between R-mode and nonlinear harmonic
excitation causes both to contribute to the measured mode intensity.

\section*{Conclusion}

Through 2.5D particle-in-cell simulations, we have investigated the excitation
of electrostatic waves in heliospheric plasmas and their
subsequent coupling to nonlinearly create electromagnetic waves, known from type
II radio bursts, focusing on the question whether the emission process of these bursts 
is significantly affected by the emission regions' background magnetic field strength.

Results show that the electrostatic wave excitation is unaffected by the strong
changes in plasma magnetisation, and creates self-consistent wave populations at
all studied ratios of electron cyclotron frequency to plasma frequency.
The subsequent nonlinear coupling of electrostatic modes causes emission
features to appear at both the fundamental ($\omega_p$) and harmonic ($2
\omega_p$) frequency, these processes, too, occur irrespective of
magnetization. Even though the
electromagnetic modes' dispersion behaviour changes drastically at higher
magnetizations, the excitation process can thus consistently drive radio
emissions with the same spectral properties and lead to the observed
continuous emission of type II radio bursts during propagation of coronal
shocks over a wide range of solar radii.

The same emission process may therefore be responsible for other heliospheric radio
emission phenomena, such as broadband type II or type IV bursts, whose assumed
emission regions may contain counterstreaming electron populations at much
higher magnetic fields. It is also a viable emission mechanism for highly
magnetized flare stars.

In a direct comparison of the development of fundamental and harmonic emission
intensity it became apparent that the harmonic emission was significantly
enhanced in the case where the electron cyclotron frequency was approximately
resonant. In which way these wave resonances may have a quantitative effect on
actual radio observations will have to be determined in a future study.

\section*{Acknowledgements}
The Academy of Finland is acknowledged for financial support (Project 133723).
The simulations for this research have been made possible through computing
grants by the Juelich Supercomputing Centre (JSC) and the CSC - IT Center for
Science Ltd., Espoo, Finland.  This work has been supported by the European
Framework Programme 7 Grant Agreement SEPServer - 262773

\bibliographystyle{aa}
\bibliography{ursg}

\begin{thebibliography}{21}
\expandafter\ifx\csname natexlab\endcsname\relax\def\natexlab#1{#1}\fi

\bibitem[{Cane \& Erickson(2005)}]{caneErickson}
Cane, H.~V. \& Erickson, W.~C. 2005, The Astrophysical Journal, 623, 1180

\bibitem[{Forbes {et~al.}(2006)Forbes, Linker, Chen, Cid, Kóta, Lee, Mann,
  Mikic, Potgieter, Schmidt, Siscoe, Vainio, Antiochos, \&
  Riley}]{WorkingGroupD}
Forbes, T., Linker, J., Chen, J., {et~al.} 2006, Space Science Reviews, 123,
  251

\bibitem[{Ganse {et~al.}(2012{\natexlab{a}})Ganse, Kilian, Spanier, \&
  Vainio}]{GanseApj2012}
Ganse, U., Kilian, P., Spanier, F., \& Vainio, R. 2012{\natexlab{a}}, \apj,
  751, 145(6pp)

\bibitem[{Ganse {et~al.}(2012{\natexlab{b}})Ganse, Kilian, Vainio, \&
  Spanier}]{ursSolarPhysics}
Ganse, U., Kilian, P., Vainio, R., \& Spanier, F. 2012{\natexlab{b}}, Solar
  Physics, 280, 551

\bibitem[{{H. Aurass} {et~al.}(2002){H. Aurass}, {B. Vrsnak}, \& {G.
  Mann}}]{TypeIIFlare}
{H. Aurass}, {B. Vrsnak}, \& {G. Mann}. 2002, A\&A, 384, 273

\bibitem[{{Hockney} \& {Eastwood}(1988)}]{HockneyEastwood}
{Hockney}, R.~W. \& {Eastwood}, J.~W. 1988, {Computer simulation using
  particles} (Bristol: Hilger, 1988)

\bibitem[{Karlick{\'y} \& Barta(2010)}]{KarlickyIAUS}
Karlick{\'y}, M. \& Barta, M. 2010, Proceedings of the International
  Astronomical Union, 6, 252

\bibitem[{Kilian {et~al.}(2012)Kilian, Burkart, \& Spanier}]{acronym11}
Kilian, P., Burkart, T., \& Spanier, F. 2012, in High Performance Computing in
  Science and Engineering '11, ed. W.~E. Nagel, D.~B. Kröner, \& M.~M. Resch
  (Berlin Heidelberg: Springer), 5--13

\bibitem[{Knock {et~al.}(2003)Knock, Cairns, \& Robinson}]{knockShockRipples}
Knock, S.~A., Cairns, I.~H., \& Robinson, P.~A. 2003, J. Geophys. Res., 108,
  1361

\bibitem[{{Knock} {et~al.}(2001){Knock}, {Cairns}, {Robinson}, \&
  {Kuncic}}]{KnockModel}
{Knock}, S.~A., {Cairns}, I.~H., {Robinson}, P.~A., \& {Kuncic}, Z. 2001,
  Journal of Geophysical Research, 106, 25041

\bibitem[{Liu {et~al.}(2009)Liu, Luhmann, Bale, , \& Lin}]{StereoData}
Liu, Y., Luhmann, J.~G., Bale, S.~D., , \& Lin, R.~P. 2009, The Astrophysical
  Journal Letters, 691, L151

\bibitem[{{Melrose}(1970)}]{Melrose1970}
{Melrose}, D.~B. 1970, Australian Journal of Physics, 23, 871

\bibitem[{{Melrose}(1986)}]{Melrose}
{Melrose}, D.~B. 1986, {Instabilities in Space and Laboratory Plasmas}
  (Instabilities in Space and Laboratory Plasmas, by D.~B.~Melrose,
  pp.~288.~ISBN 0521305411.~Cambridge, UK: Cambridge University Press)

\bibitem[{Mikhailovskii(1981)}]{MikhailovskiiCherenkov}
Mikhailovskii, A.~B. 1981, Plasma Physics, 23, 413

\bibitem[{{Nelson} \& {Melrose}(1985)}]{NelsonMelrose}
{Nelson}, G.~J. \& {Melrose}, D.~B. 1985, {Type II bursts} (Cambridge
  University Press), 333--359

\bibitem[{{Osten} \& {Bastian}(2008)}]{AreciboObservation}
{Osten}, R.~A. \& {Bastian}, T.~S. 2008, \apj, 674, 1078

\bibitem[{Pohjolainen {et~al.}(2013)Pohjolainen, Allawi, \&
  Valtonen}]{PohjolainenWideband}
Pohjolainen, S., Allawi, H., \& Valtonen, E. 2013, A\&A, 558, A7

\bibitem[{Pulupa \& Bale(2008)}]{Pulupa2007}
Pulupa, M. \& Bale, S.~D. 2008, The Astrophysical Journal, 676, 1330

\bibitem[{Spanier \& Vainio(2009)}]{SpanierVainio09}
Spanier, F. \& Vainio, R. 2009, Advanced Science Letters, 2, 337

\bibitem[{Wild \& McCready(1950)}]{WildMcCready}
Wild, J. \& McCready, L. 1950, Australian Journal of Scientific Research, 3,
  387

\bibitem[{{Willes} \& {Cairns}(2000)}]{WillenGeneralizedLangmuir}
{Willes}, A.~J. \& {Cairns}, I.~H. 2000, Physics of Plasmas, 7, 3167

\end{thebibliography}

\end{document}